\pdfoutput=1
\documentclass[sigplan,10pt]{acmart}

\usepackage{makecell}
\usepackage{subfig}
\usepackage{multirow,tabularx}
\usepackage{balance}

\usepackage{amsfonts}
\usepackage{booktabs}
\usepackage{siunitx}
\usepackage{xspace}

\newcommand{\cm}{{Dirigent}\xspace}
\newcommand{\knative}{{Knative}\xspace}

\newcommand{\anaedit}[1]{{\textcolor{black}{#1}}}

\newcommand{\lazar}[1]{{\textcolor{black}{#1}}}

\newcolumntype{b}{X}
\newcolumntype{s}{>{\centering{\hsize=.5\hsize}}X}
\newcolumntype{x}{>{\hsize=.25\hsize}X}
\newcolumntype{y}{>{\hsize=.5\hsize}X}

\setcopyright{rightsretained}
\begin{document}

\date{}

\title{Dirigent: Lightweight Serverless Orchestration}


\author{Lazar Cvetkovi\'{c}}
\affiliation{%
  \institution{ETH Zurich}
  \country{}
}
\email{lazar.cvetkovic@inf.ethz.ch}

\author{François Costa}
\affiliation{%
  \institution{ETH Zurich}
  \country{}
}
\email{francois.costa@inf.ethz.ch}

\author{Mihajlo Djokic}
\affiliation{%
  \institution{ETH Zurich and IBM Research Europe}
  \country{}
}
\email{djokicm@ethz.ch}

\author{Michal Friedman}
\affiliation{%
  \institution{ETH Zurich}
  \country{}
}
\email{michal.friedman@inf.ethz.ch}

\author{Ana Klimovic}
\affiliation{%
  \institution{ETH Zurich}
  \country{}
}
\email{aklimovic@ethz.ch}

\acmYear{2024}\copyrightyear{2024}

\acmConference[SOSP '24]{ACM SIGOPS 30th Symposium on Operating Systems Principles}{November 4--6, 2024}{Austin, TX, USA}
\acmBooktitle{ACM SIGOPS 30th Symposium on Operating Systems Principles (SOSP '24), November 4--6, 2024, Austin, TX, USA}
\acmDOI{10.1145/3694715.3695966}
\acmISBN{979-8-4007-1251-7/24/11}




\settopmatter{printfolios=false}
\maketitle
\pagestyle{plain}

\subsection*{Abstract}
While Function as a Service (FaaS) platforms can initialize function sandboxes on worker nodes in 10-100s of milliseconds, the latency to schedule functions in real FaaS clusters can be orders of magnitude higher. The current approach of building FaaS cluster managers on top of legacy orchestration systems (e.g., Kubernetes) leads to high scheduling delays when clusters experience high sandbox churn, which is common for FaaS.
Generic cluster managers use many hierarchical abstractions and internal components to manage and reconcile cluster state with frequent persistent updates. This becomes a bottleneck for FaaS since the cluster state frequently changes as sandboxes are created on the critical path of requests.
Based on our root cause analysis of performance issues in existing FaaS cluster managers, we propose \textit{\cm}, a clean-slate system architecture for FaaS orchestration with three key principles. First, \cm optimizes internal cluster manager abstractions to simplify state management. Second, it eliminates persistent state updates on the critical path of function invocations, leveraging the fact that FaaS abstracts sandbox locations from users to relax exact state reconstruction guarantees. Finally, \cm runs monolithic control and data planes to minimize internal communication overheads and maximize throughput. We compare \cm to state-of-the-art FaaS platforms and show that \cm reduces 99\textsuperscript{th} percentile per-function scheduling latency for a production workload by 2.79$\times$ compared to AWS Lambda. \cm can spin up 2500 sandboxes per second at low latency, which is 1250$\times$ more than Knative.




\section{Introduction}\label{sec:intro}

Serverless computing --- in particular, Function as a Service (FaaS) --- is an appealing paradigm of cloud computing as it raises the user's level of abstraction to the cloud and alleviates users from the burden of explicitly managing server resources~\cite{serverless_cacm}. In addition to ease of use, to be practical, a FaaS platform must execute functions in securely isolated environments (i.e., sandboxes) while minimizing end-to-end latency and maximizing function execution throughput per machine for cost-efficiency~\cite{agache:firecracker}.

While initializing function sandboxes on \textit{worker nodes} takes 10-100s of milliseconds\footnote{We assume that function container images are cached on worker nodes.} with today's FaaS worker system software~\cite{agache:firecracker, young:gvisor, du:catalyzer, brooker:firecracker_snapshots, particle, oakes:sock, virtines}, we find that the \textit{end-to-end} latency to initialize function sandboxes is often one or more orders of magnitude higher in operational FaaS environments. This is because initialization involves more than creating and starting a sandbox on a worker node. First, the cluster manager receiving function invocations must schedule the sandbox to be created on a particular worker node. Then, after the sandbox is ready, the cluster manager must plug it into the cluster so that it starts receiving traffic. While scheduling a single sandbox at a time can be relatively quick, we find that scheduling delay dominates  when the cluster manager concurrently schedules many sandboxes.

\begin{figure}
    \centering
    \includegraphics[trim={0 0.6cm 0 0},width=\linewidth]{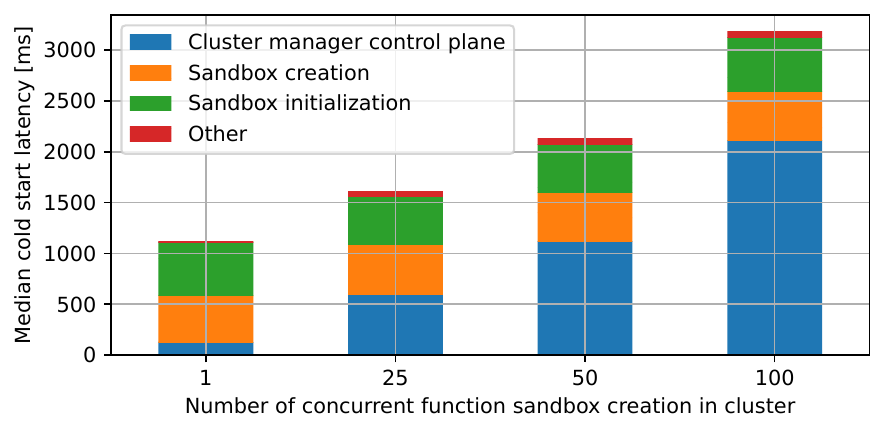}
    \caption{End-to-end latency breakdown of cold invocation bursts in \knative. Sandbox creation involves sequentially creating two containers: user-code container and its sidecar. Sandbox init is the time it takes to pass health probes.} 
    \label{fig:concurrent_sandbox_creation_sweep}
\end{figure}

\begin{figure}
    \centering
    \includegraphics[trim={0 0.75cm 0 0},width=\linewidth]{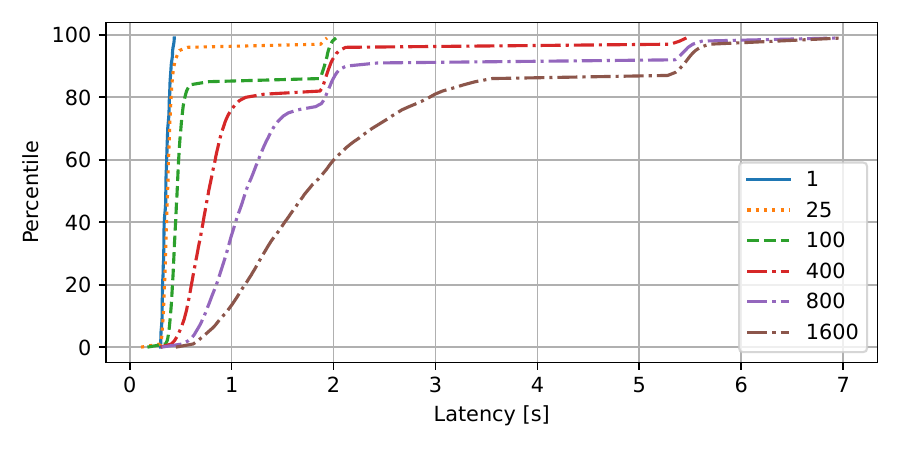}
    \caption{AWS Lambda end-to-end latency CDFs with different cold start bursts of hello-world functions. We pre-cache container images, based on insights from Brooker et al.~\cite{brooker:firecracker_snapshots}.} 
    \label{fig:aws-burst}
\end{figure}

Figure~\ref{fig:concurrent_sandbox_creation_sweep} shows how the end-to-end function initialization latency --- and in particular the latency contribution of the cluster manager --- scales as we vary the number of concurrent sandbox creations in the \knative Serving\footnote{We refer to Knative Serving simply as Knative from now on.}~\cite{knative} FaaS platform. 
The cluster manager adds 2 seconds of delay when it concurrently schedules 100 sandboxes in a burst. In Figure~\ref{fig:aws-burst}, we perform a similar experiment on AWS Lambda~\cite{aws:lambda}. While we cannot measure the cluster manager component of latency for proprietary FaaS platforms, Figure~\ref{fig:aws-burst} confirms that the same symptoms are present: end-to-end latency increases as we scale concurrent cold starts. 
This is problematic because multi-tenant, production FaaS workloads~\cite{shahrad:serverless} require over 300 sandbox creations per second on average, with bursts as high as 8000 (see \S\ref{background:requirements}), as FaaS applications consist of many short-lived, sporadically invoked functions~\cite{wang:faasnet, joosen:how_does_it_function}.

So where does FaaS scheduling overhead come from and what can we do about it? \anaedit{Although serverless scheduling research has focused on scheduling \textit{policies}~\cite{singhvi:atoll, roy:icebreaker, mittal:Mu, fuerst:FaasCache, kaffes:hermod, palette}, we find that the \textit{mechanisms for propagating policy decisions} from the cluster manager to worker nodes are a bottleneck. We identify high software bloat from the current approach of building FaaS cluster managers on top of legacy orchestration systems that were originally designed to manage long-lived, stateful datacenter applications. 
In particular, many FaaS cluster managers~\cite{openfaas, openwhisk, fission, kubeless} rely on Kubernetes (K8s)~\cite{kubernetes} to deploy sandboxes on worker nodes, monitor and manage cluster state, and recover from component failures. While K8s provides useful functionality, it leads to high scheduling latency and limits scheduling throughput in high-churn environments like FaaS (i.e.,  when sandboxes need to be frequently created and destroyed).}

For example, we take Knative~\cite{knative} as a representative FaaS cluster manager. It is used in open-source FaaS frameworks like vHive~\cite{ustiugov:benchmarking} and in Google's commercial FaaS offering~\cite{google:cloud_run}. 
Knative builds on K8s, adding invocation-based autoscaling, such that sandboxes can scale (potentially down to zero) for each function based on its invocations. It uses the K8s API to represent a sandbox as a Pod with a Service Endpoint, 
belonging to a ReplicaSet, managed as a Deployment. Under the hood, K8s runs a separate controller to manage the state associated with each of these abstractions.  Each controller periodically executes a state reconciliation loop~\cite{sieve}, which involves watching for updates and writing updates to a strongly consistent persistent database. 
Hence, creating a single sandbox involves 10s of RPCs and sequential database updates in the cluster manager. With high sandbox churn in FaaS clusters, long queuing delays arise, as seen in Figure~\ref{fig:concurrent_sandbox_creation_sweep}.

\anaedit{While one could try to retrofit K8s to improve its sandbox scheduling performance, we derive design principles for FaaS cluster management that fundamentally diverge from the K8s system design philosophy and hence opt for a clean-slate cluster manager design.} We propose \textit{\cm}, a new system architecture for cluster management, specialized for FaaS. 
\cm exposes the same user API as current FaaS platforms (i.e., users register and invoke functions). Instead of relying on a generic system like K8s to orchestrate sandboxes, 
\anaedit{\cm leverages the unique characteristics of FaaS to optimize scheduling throughput and latency.} 

We design \cm with three key principles. First, \cm \textit{simplifies cluster management abstractions} to minimize the volume and complexity 
of the cluster state. 
Compared to orchestrators that expose a variety of hierarchical abstractions (e.g., ReplicaSets and Deployments in K8s) to support declaratively grouping, scaling, and restarting sandboxes, \cm uses a lean set of abstractions designed for managing sandboxes of stateless, independent serverless functions. 
Second, \cm \textit{does not modify any persistent cluster state on the critical path of function invocations}. In particular, when \cm needs to create new sandboxes for an incoming function invocations (i.e., ``cold starts''), it does not  persist cluster state about the number and location of sandboxes for each function; it only maintains this state in memory.
This means that, in contrast to traditional cluster managers, \cm may not always restore the cluster to an identical state when a control plane component replica fails.  However, relaxing the exact recovery of sandboxes is suitable for FaaS as the cluster manager abstracts sandbox information from end-users and continuously autoscales sandboxes to match the current invocation load. 
Finally, \cm redesigns the cluster manager system architecture with a \textit{monolithic control plane} to minimize RPC overheads and a \textit{monolithic data plane} to reduce hops on the critical path of warm invocations. 

We show that Dirigent supports 2500 cold starts per second, which is 1250$\times$ more than Knative. For the Azure Functions trace,
Dirigent reduces per-function scheduling latency at the 99\textsuperscript{th} percentile by 403$\times$ compared to Knative and 2.79$\times$ compared to AWS Lambda. Dirigent provides the same fault tolerance guarantees for FaaS users while enabling faster recovery times from control plane, data plane, and worker node failures. 
\lazar{Dirigent is an open-source project available at: \url{https://github.com/eth-easl/dirigent}.}

\section{Background and Motivation}\label{sec:background}

We outline the requirements for FaaS cluster management (\S\ref{background:requirements}) and analyze the fundamental mismatch between these requirements and K8s, which is used today as the foundation in many FaaS platforms (\S\ref{background:k8s-mismatch}). We discuss alternative cluster managers and why they are not suitable for FaaS (\S\ref{background:alternatives}).

\subsection{FaaS Cluster Management Requirements}
\label{background:requirements}

A FaaS cluster manager manages function registrations and schedules function invocations for execution on worker nodes. Scheduling in the context of FaaS involves three aspects: \textit{autoscaling} (i.e., creating and tearing down) sandboxes per function based on invocations, \textit{placing} sandboxes across workers to optimize performance and resource efficiency, and \textit{load-balancing} invocations across sandboxes. A FaaS cluster manager is also responsible for keeping the FaaS service operational despite potential cluster component failures. 
We summarize the key requirements for a FaaS cluster manager and discuss the associated challenges that arise due to the unique characteristics of FaaS workloads:

\begin{figure}
    \centering
    \includegraphics[trim={0 1cm 0 0},width=\linewidth]{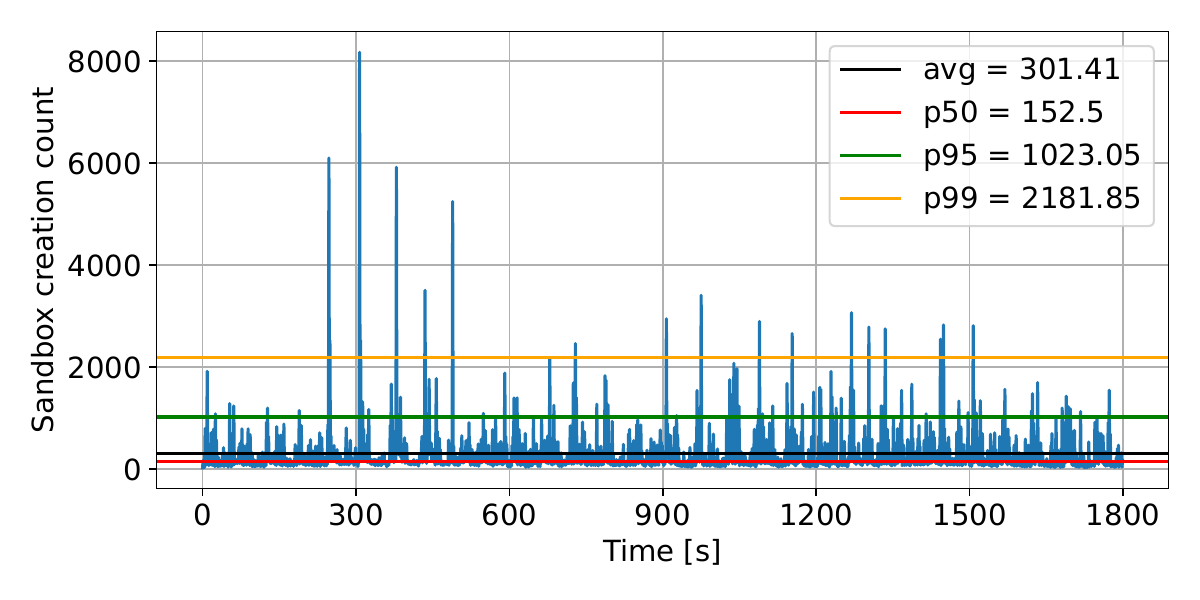}
    \caption{Rate of sandbox creation over time in a 30-minute window (after 10-min warmup) of the 70K function Azure trace~\cite{shahrad:serverless}, simulated on a 1000 worker-node cluster with default \knative scheduling policies. Each sandbox processes 1 request at a time, the default for FaaS platforms~\cite{aws:sandbox_concurrency,gcf:invocation_level_guarantees}.}
    \label{fig:invocation-load}
\end{figure}

\textbf{R1) High throughput scheduling.}
A FaaS cluster manager must be able to create, place, and tear down function sandboxes and load-balance incoming requests with high throughput.
FaaS workloads involve bursty and unpredictable function invocations~\cite{shahrad:serverless}. Keeping many warm sandboxes available in DRAM is expensive, so the cluster manager must frequently create and destroy function sandboxes. 
Figure~\ref{fig:invocation-load} plots the number of sandbox creations in the Azure trace over a 30-minute time window when simulating the trace on a 1000-node cluster with the default autoscaling, load-balancing, and placement policy in Knative~\cite{knative:autoscaling_policy,knative:load_balancing_policy,kubernetes:placement_policy}. 
For this workload, the cluster manager creates 300 sandboxes per second on average, with bursts of thousands of sandboxes per second. Even if we configure the scaling policy to have infinite keep-alive (i.e., never downscale functions sandboxes after an invocation), the cluster manager still needs to create 229 sandboxes per second on average and 1551 per second at the 99\textsuperscript{th} percentile. This is due to inevitable cold starts when functions are invoked for the first time. 
In contrast, traditional cluster managers 
do not optimize sandbox creation and placement throughput, since sandboxes are often pre-deployed off the critical path of requests and sandbox creation is amortized for traditional, long-lived applications. 

\textbf{R2) Low latency scheduling.}
Since serverless functions are often short-lived (e.g., 50\% of functions in the Azure Functions trace~\cite{shahrad:serverless} execute within a second), the cluster manager must schedule functions with low latency (ideally less than tens of ms) on the critical path.

\textbf{R3) Fault tolerance.}
We distinguish between component-level and request-level fault tolerance. The FaaS cluster manager must provide \textit{component-level} fault tolerance, i.e., ensure the platform remains operational and able to serve new invocations despite worker, data plane, or control plane node failures. The platform should minimize the impact of component failures on the end-to-end invocation latency.

\textit{Request-level} fault tolerance concerns requests that are \textit{in-flight} in the cluster when a failure occurs. 
Though desirable~\cite{boki, beldi, halfmoon}, existing FaaS platforms generally do not provide request-level fault tolerance. For synchronous invocations --- where the client blocks until receiving a response --- state-of-the-art FaaS platforms rely on users to re-invoke a function ~\cite{azure-retry,aws:invocation_level_guarantees,gcf:invocation_level_guarantees} in case an invocation is lost (e.g., if a worker node fails in the middle of execution). 
Some FaaS platforms, like AWS Lambda, also support asynchronous invocations with a persistent queue that buffers invocations and can retry invocations in case of timeouts to provide at-least-once invocation guarantees. Since a function may get invoked (and partially executed) more than once, FaaS platforms advise users to write idempotent functions~\cite{azure-idempotent, aws-idempotent}.

\textbf{Non-requirements.} 
A FaaS cluster manager does not expose the exact number and location of sandboxes to end-users, nor it needs to support direct communication between sandboxes~\cite{pocket}. Hence, in case a particular sandbox fails, it is not necessary to restore the cluster to an identical state. Redeploying sandboxes is acceptable and straightforward as FaaS functions are independent and stateless, in contrast to generic applications which may have complex workflow chains and whose components spread across different sandboxes may have complex inter-communication patterns. 

\subsection{The Kubernetes -- FaaS Mismatch}
\label{background:k8s-mismatch}

We now discuss the mismatch between the FaaS cluster manager requirements in \S\ref{background:requirements} and K8s-based cluster managers, which are common in current FaaS platforms\lazar{~\cite{cncf:survey_2020}}, such as Knative~\cite{knative}, OpenWhisk\footnote{\lazar{OpenWhisk can run in a non-K8s Docker setup for clusters with less than 10 nodes and 100 containers~\cite{docker:swarm:what_is_it}, but the K8s deployment is encouraged~\cite{openwhisk:k8s_recommendation}.}}~\cite{openwhisk}, OpenFaaS~\cite{openfaas}, Fission~\cite{fission}, Kubeless~\cite{kubeless}, Cloudburst~\cite{sreekanti:cloudburst}, and Google Cloud Run for Anthos~\cite{google:cloud_run}.  
The K8s-based cluster managers in these platforms ensure component-level fault tolerance for FaaS (R3 in \S\ref{background:requirements}). However, we find that building on generic K8s API abstractions and inheriting the microservice-based architecture of K8s makes cluster managers unfit for high throughput and low latency FaaS workload scheduling (R1 and R2). 

\begin{figure}
    \centering
    \includegraphics[trim={0 0.25cm 0 0},width=\linewidth]{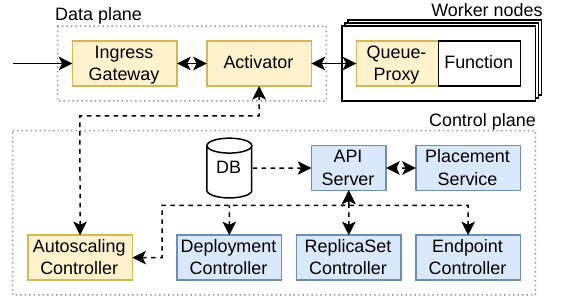}
    \caption{Knative system architecture, which builds on K8s. This diagram is simplified, showing only key components which all run as independent microservices. K8s components are blue, while yellow components are added by Knative.}
    \label{fig:k8s_scaling_diagram}
\end{figure}

\textbf{\knative case study:} We take \knative~\cite{knative} as a representative FaaS cluster manager, as it is open-source and widely used~\cite{cncf:survey_2020}, both in research~\cite{ustiugov:benchmarking} and commercially~\cite{google:cloud_run}.
Figure~\ref{fig:k8s_scaling_diagram} shows the Knative system architecture and how it builds on K8s components and concepts. 
The K8s API~\cite{kubernetes:api_specification} provides concepts, such as Deployments, ReplicaSets, and Endpoints, which can be used to monitor and control cluster state at different levels of abstraction. For example, a Pod (the minimal scheduling unit in K8s) can be horizontally scaled as a ReplicaSet, 
a low-level K8s object that ensures a specified number of replicas are running at all times. K8s can manage ReplicaSets with a higher-level object, a Deployment, which provides additional features like rolling updates and rollbacks. 
K8s stores state for all objects in the cluster in a strongly-consistent database. K8s also implements multiple controllers that run reconciliation loops for objects like Deployments and ReplicaSets to converge the actual system state to the desired state. 
To use K8s as the underlying resource orchestrator for FaaS, Knative extends K8s with an additional set of controllers to implement invocation-based autoscaling. The Knative autoscaling controller supports scaling a function to zero sandboxes at low load. This is necessary for FaaS as the default K8s Horizontal Pod Autoscaler cannot scale a function to zero, i.e., has no support for cold starts but scales sandboxes based on generic metrics like CPU and memory utilization~\cite{kubernetes:hpa}. Knative also adds a component to buffer requests for cold starts (\textit{Activator}) and a per-Pod sidecar component (\textit{Queue-Proxy}) to throttle the number of concurrent requests each Pod can process. 

\begin{figure}
    \centering
    \includegraphics[trim={0 0.5cm 0 0},width=\linewidth]{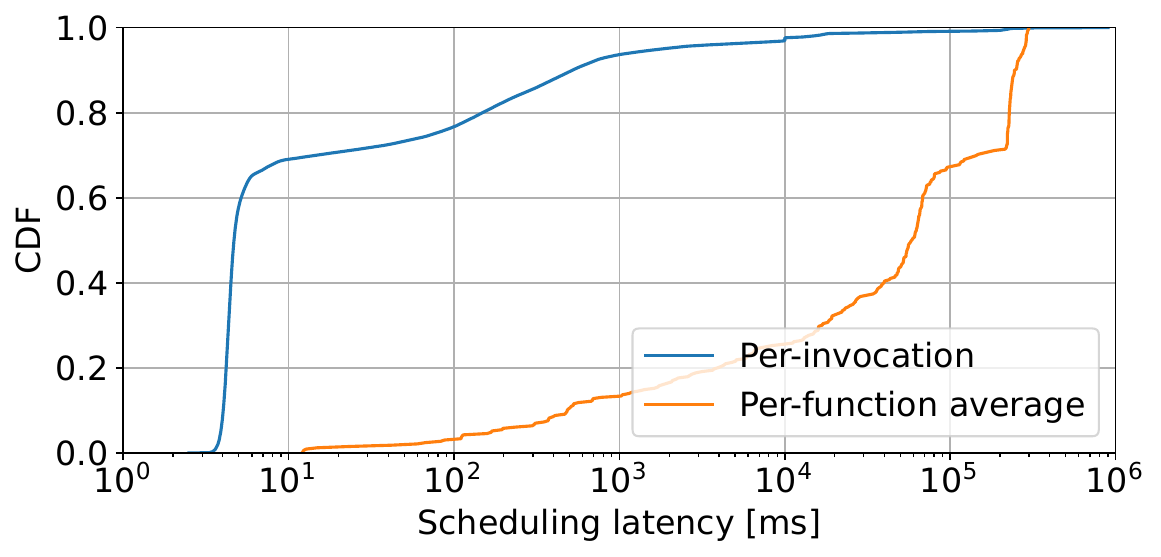}
    \caption{CDF of per-invocation scheduling latency and per-function mean scheduling latency when executing 500-function Azure trace~\cite{ustiugov:in_vitro, shahrad:serverless} on a 93-worker cluster.}
    \label{fig:motivation_cluster_manager_latency_cdf}
    \vspace{-5pt}
\end{figure}

While the K8s API provides convenient abstractions and the K8s architecture is modular and extensible, we find that implementing a FaaS cluster manager on top of K8s has high performance overhead. For example, Figure~\ref{fig:motivation_cluster_manager_latency_cdf} shows the cumulative distribution of \knative scheduling latency when running a 500-function sample of the Azure production trace~\cite{shahrad:serverless} on a 93 worker-node cluster. Scheduling has long tail latency. One third of functions experience an average scheduling latency greater than or equal to 100 seconds, whereas many functions only execute for milliseconds. 

To understand the root cause of this latency overhead, we analyze which function invocations experience high scheduling delays. We find it is functions invoked while the cluster manager is orchestrating a large number of concurrent sandbox creations. Figure~\ref{fig:concurrent_sandbox_creation_sweep} confirms \knative cluster manager latency increases significantly when the cluster experiences multiple concurrent cold starts. We validate our findings by running cold start invocation microbenchmarks on Google Cloud Run for Anthos~\cite{google:cloud_run}, a commercial \knative offering. We see similar latency patterns as we scale cold start invocations. 

The fundamental bottleneck is the complex critical path of sandbox creation in \knative, as the system relies on multiple K8s-based controllers to reconcile desired and actual cluster state. While computing the desired state (i.e., executing the autoscaling and placement algorithms) is fast, reconciling the cluster state is highly inefficient for several reasons. First, by design, K8s components cannot exchange information directly, even if they run in the same process. The K8s controllers can only exchange information through synchronous read-modify-write sequences to a centralized cluster state database, etcd~\cite{etcd}. Hence, creating a new sandbox in the cluster involves multiple RPCs between controllers and the database front-end (the API server). These operations are not commutative and hence impede scalability~\cite{scalable-commutativity-rule}. Second, the volume of state exchanged in RPC calls is large as K8s manages state with key-value pairs that average 17kB in size in our experiments and are represented as deeply nested trees. As a result, we find the API server spends significant CPU cycles on data serialization. When invoking cold starts at a steady rate, we find \knative can only support 2 cold starts per second before scheduling latency saturates (see Figure~\ref{fig:system_performance_comparison}) due to the API server saturating CPU resources. Finally, K8s serializes and persists cluster state updates with strong consistency. While serializing and persisting updates enables restoring the cluster state to an identical state as before a failure occurred, it limits sandbox creation throughput.

A natural approach to scale sandbox scheduling throughput is to deploy functions across independent sub-clusters.
However, supporting the median sandbox creation rate in the Azure trace simulation shown in Figure~\ref{fig:invocation-load} (152~sandbox creations per second) would require spreading invocations across $\sim$90 separate sub-clusters, each managed by a separate \knative instance. 
\lazar{Each sub-cluster would require separate nodes for control and data plane replicas. An additional load-balancing layer would add an extra hop for all requests. Furthermore, the division of the cluster into sub-clusters would reduce global visibility of the load across machines, which can degrade scheduling decision quality~\cite{yaq}.} 

\textbf{Generalizing beyond \knative:} To test if our findings generalize to other K8s-based cluster managers besides Knative, we experimented with OpenWhisk~\cite{openwhisk}.
Also, we tested bypassing K8s abstractions, such as Deployments and ReplicaSets, and instead directly created and managed Pods. In both cases, we observe high cold start latency with concurrent cold starts, confirming that even creating and tearing down the minimal type of K8s objects (Pods) has high overhead at the high churn rate required by FaaS applications. 

We also observe that increasing concurrent sandbox creations significantly impacts AWS Lambda cold start latency (Figure~\ref{fig:aws-burst}), however, we do not have access to the platform's cluster manager implementation to analyze the root cause.

\subsection{Related Work}

\label{background:alternatives}
\label{seb:background:related_work}

\textbf{Alternative cluster managers.} Cluster manager design is an active research area, with many alternatives to K8s~\cite{borg_omega_k8s, schwarzkopf:omega}. 
However, data center cluster managers~\cite{verma:borg_2015,apollo,twine, hindman:mesos,vavilapalli:yarn,openstack,isard:quincy,gog:firmament,ousterhout:sparrow,quasar, paragon, tarcil,karanasos:mercury} are typically designed to orchestrate long-living applications. For such applications, sandbox creation is amortized and not on the critical paths of requests. FaaS, in contrast, has much shorter sandbox lifetimes and higher churn. 
To orchestrate thousands of nodes and applications, systems such as Mesos and YARN~\cite{hindman:mesos, vavilapalli:yarn} embed all inter-component communication into periodic heartbeats. 
However, long heartbeat periods lead to poor responsiveness, which is not suitable for FaaS workloads.
Quincy~\cite{isard:quincy} and Firmament~\cite{gog:firmament} focus on scheduling policy design and explore the tradeoff of computational efficiency vs. decision quality, but ignore how the cluster manager system architecture affects decision propagation speed in the cluster. Sparrow~\cite{ousterhout:sparrow} improves scalability by decentralizing scheduling, however, trading off global knowledge of the load on each worker node can degrade decision quality~\cite{yaq}. 

Many prior works explore complementary, such as reducing interference between the co-located workloads~\cite{quasar, paragon, tarcil}.
Mercury~\cite{karanasos:mercury} explores tradeoffs for collocating long-running analytic jobs with latency-critical workload. Omega~\cite{schwarzkopf:omega} explores tradeoffs between centralized and distributed scheduler designs.
DCM~\cite{suresh:declarative_cluster_manager} proposes a new cluster management architecture to simplify scheduling policy implementation and debugging for developers, by enabling declarative SQL queries to a relational cluster state database. 
Sieve~\cite{sieve} and Anvil~\cite{anvil} address correctness challenges with state reconciliation systems like K8s to improve reliability.

\textbf{Cluster management for FaaS.} 
The closest to our work is a study characterizing the gap between FaaS research and real-world systems, which also identifies high cold start latency when scheduling many sandboxes~\cite{huawei:serverless_gap}. We analyze the root-cause and design \cm to alleviate cluster manager control plane bottlenecks.  Ilúvatar~\cite{fuerst:iluvatar} is complementary work that reduces warm start scheduling overheads originating on worker nodes. 
Most work on FaaS orchestration has focused on autoscaling, load-balancing, and placement policies to reduce the frequency and overhead of cold starts, improve end-to-end performance, and resource efficiency~\cite{roy:icebreaker,mittal:Mu,kaffes:hermod,singhvi:atoll,shahrad:serverless,palette,fuerst:Locality-Aware-Load-Balancing,ousterhout:sparrow}. 
These works build on top of existing FaaS cluster manager system architectures, in which the state management performance bottlenecks described in \S\ref{background:k8s-mismatch} remain. 

\textbf{Adapting K8s.} Some works have adapted K8s for different use cases. KOLE~\cite{zhang:kole} adapts K8s for the edge environment and manages to scale K8s to 1M nodes but at the expense of abolishing dynamic Pod creation and scheduling, which is not suitable for FaaS. K3s~\cite{k3s} is a lightweight K8s for IoT and edge environments. Although the single-process version of K8s is easy to deploy, we observed the system suffers from many of the same performance issues as the baseline K8s. Faasd ~\cite{faasd} targets single-node resource-constrained edge setups, while we target FaaS cloud clusters.

\section{\cm Design Approach}
\label{sec:system_design}

To address the scheduling overheads in state-of-the-art FaaS platforms, we propose \textit{\cm}, a new cluster manager catered for FaaS. 
\cm maintains the same serverless end-user API as today's FaaS platforms (i.e., users register and invoke functions) such that applications designed for AWS Lambda or \knative can seamlessly run on \cm.
To meet the performance and fault tolerance requirements of FaaS applications (discussed in \S\ref{background:requirements}), we derive system design principles based on insights from our analysis of K8s-based FaaS systems, summarized in Table~\ref{table:k8s_lessons_dirigent_principles}. 
This results in a clean-slate system design, which we present below.

\begin{table*}[t]
  \begin{center}
    \centering
    \begin{tabularx}{\linewidth}{y|y}
         \hline
         \textbf{Feature of K8s-based FaaS system design that contributes to high scheduling latency} & \multicolumn{1}{c}{\textbf{Insight for \cm design}} \\ 
         \hline 
         \hline
         
         Managing a large volume of state for many, hierarchical abstractions in K8s (e.g., Deployments, ReplicaSets). & Simple internal cluster management abstractions. \\
         \hline

         Persisting and serializing each cluster state update on the critical path of cold function invocations. & Persistence-free latency-critical operations, relaxing exact cluster state reconstruction as it is abstracted from FaaS users. \\
         \hline

         Microservice-based control plane with RPC communication between components. & Monolithic control plane. \\
         \hline

         Per-sandbox sidecars on workers for concurrency throttling. & Monolithic data plane for request throttling.\\
         \hline
         
         \hline
    \end{tabularx}
    \caption{
        \cm's design principles, based on insights from our performance issues analysis in K8s-based FaaS systems.
    }
    \label{table:k8s_lessons_dirigent_principles}
  \end{center}
\end{table*}

\subsection{System Overview}\label{design:overview}

\begin{figure}
    \centering
    \includegraphics[trim={0.3cm 0.5cm 0 0},width=1.05\linewidth]{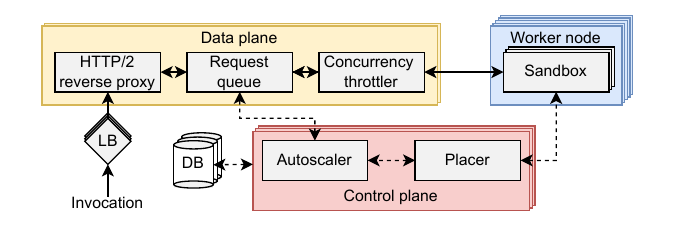}
    \caption{System diagram of \cm cluster manager.}
    \label{fig:dirigent_system_diagram}
\end{figure}

\textbf{System architecture.} Figure \ref{fig:dirigent_system_diagram} shows \cm's system architecture. \lazar{Each Dirigent component runs as an independent process, preferably on a separate physical machine, and can be replicated independently.} The \emph{control plane} is responsible for monitoring cluster components, autoscaling, placing sandboxes on worker nodes, and persisting cluster state. 
\lazar{Only one control plane component replica is active at a time. The active control plane component replica persists some of its state to a database, which is replicated across nodes with strong consistency.}  
The \emph{data plane} load balances incoming invocations to worker nodes, buffers invocations waiting for a sandbox, and limits the number of requests that a sandbox processes in parallel (concurrency throttling). 
\lazar{Data plane component replicas are all active and independent.}
The front-end \emph{load balancer} (LB in Figure~\ref{fig:dirigent_system_diagram}) spreads incoming invocations across data plane components.
\emph{Worker nodes} execute function invocations and create/destroy sandboxes when instructed by the control plane. \S\ref{design:life-of-req} describes the life of a function invocation request through the system.

\textbf{\cm API.} The bold Client caller rows at the top of Table~\ref{table:dirigent_control_plane_api} show \cm's end-user API, which corresponds to FaaS platforms like AWS Lambda and Knative. \lazar{Users register functions in \cm by providing container images. Hence, workloads require no porting effort. Functions simply need to expose a gRPC/HTTP server, as in today's FaaS platforms. Users can directly invoke functions or configure triggers (e.g. timer events) to invoke functions.} The other rows in Table~\ref{table:dirigent_control_plane_api} show the internal calls supported between \cm's components to send metrics, add/remove components, and perform leader election.

\subsection{Design Principles}
\label{sec:design_principles}

\cm's design is based on principles that address the performance issues we identified in K8s-based FaaS cluster managers (Table~\ref{table:k8s_lessons_dirigent_principles}). We discuss each principle below.

\textbf{Simple internal abstractions.}
In contrast to Knative, which uses a plethora of K8s objects (e.g., Deployments, ReplicaSets, Endpoints
), \cm defines only four fundamental types of objects that the control plane orchestrates, shown in Table~\ref{table:dirigent_abstractions}. 
The \textit{Function} abstraction represents a function that a user registers with a name, container image URL, and exposed port. \cm uses this information as a recipe for creating sandboxes of that function. \cm keeps track of per-function scheduling configurations (e.g., autoscaling knobs, resource quotas, placement constraints) and monitors per-function scheduling metrics, such as the number of inflight requests of that function. The \textit{Sandbox} abstraction (analogous to K8s Pod) represents information about the sandbox state on a worker node, such as the sandbox name, IP address, port, and the name of the worker node it resides on. \textit{DataPlane} and \textit{WorkerNode} objects store the IP addresses and ports of respective components so that a control plane can re-establish communication channels in case they fail.

Minimizing the number of internal abstractions minimizes the amount of state that \cm needs to maintain, improves resource efficiency, and avoids double bookkeeping and its associated consistency overheads. Moreover, it reduces the number of state updates needed whenever the autoscaling algorithm triggers a sandbox creation or teardown. In Knative, a sandbox creation triggers updates to multiple hierarchical objects (e.g., Deployment, ReplicaSet, Endpoint, Routes) via their associated state reconciliation controllers. On the other hand, \cm updates a single Sandbox object and forwards data plane components an updated list of sandboxes. 

In addition to managing fewer objects, \cm also minimizes the state stored per object. For example, by tailoring \cm's state management for the FaaS use case, we store the sandbox state in 16 bytes, compared to a K8s Pod resource definition, which we find can be as big as 17~KB. 
We find \knative uses K8s abstractions to store large function-related metadata in YAML format as raw Unicode text. This data includes annotations and labels, environment variables, sandbox state transition timestamps, and control messages. 
The schema features many long keys, which amplify serialization overheads. 
In \cm, we adopt a minimalist metadata and storage schema and store state in a serialized binary format.

\begin{table}[t]
  \begin{center}
    \centering
    \begin{tabularx}{\linewidth}{xyx}
         \hline
         Caller & Operation & Callee \\ 
         \hline
         \hline
        
         \multirow{2}{*}{\textbf{Client}} & \textbf{(De)-Register function} & \textbf{CP} \\
         & \textbf{Invoke function} & \textbf{DP} \\
         \hline
         
         \multirow{4}{*}{\thead{Data plane\\(DP)}} 
         & (De)-Register data plane & CP \\
         & List registered functions & CP \\
         & Send scaling metric & CP \\
         & Send heartbeat & CP \\

         \hline
        
         \multirow{5}{*}{\thead{Control plane\\(CP)}} & Add/remove function & DP \\
         & Add/remove LB endpoint & DP \\
         & Create/Kill sandbox & WN \\
         & List sandboxes & WN \\
         & Vote for leader election & CP \\
         \hline
        
         \multirow{2}{*}{\thead{Worker node\\(WN)}} & (De)-Register worker & CP \\
         & Send heartbeat & CP \\
         \hline
    \end{tabularx}
    \caption{\cm API. Bold operations are exposed to users. Others are internal calls between \cm components. 
    } 
    
    \label{table:dirigent_control_plane_api}
  \end{center}
\end{table}

\textbf{Persistence-free latency-critical operations.}
\lazar{The control plane persists (with strong consistency to a replicated database) only the minimal state required for the control plane to operate correctly after recovering from a failure. Table~\ref{table:dirigent_abstractions} shows the cluster state \cm maintains in memory and the checkmarks in the last column indicate which state is persisted. 
\cm's control plane does not persist state that is recoverable from other cluster components. This includes the \textit{Sandbox} state (which can be recovered from worker nodes) and \textit{Function} scheduling metrics (which can be inferred from data plane traffic).}

\lazar{In contrast, in FaaS platforms like Knative, K8s mandates that every cluster state change (e.g., adding a Pod to a ReplicaSet) is persisted in a centralized, strongly consistent database, such that K8s can restore the cluster to the exact state as before the failure. We argue that such strong guarantees -- and the performance overheads that they come with -- are not fundamentally necessary in FaaS clusters. }

\lazar{Specializing the cluster manager design for FaaS opens opportunities for relaxing state reconstruction guarantees. For example, if a FaaS cluster fails, sandboxes can be started on different worker nodes, as IP addresses and placement information are not exposed to end-users. Additionally, the cluster need not recover the same number of sandboxes as the traffic often varies significantly over short time windows.
While relaxing state reconstruction guarantees may be unsuitable for a generic cluster manager with an arbitrary workload, \cm still satisfies the component-level fault tolerance requirements of FaaS platforms (R3 in \S\ref{background:requirements}). In \S\ref{sec:component_level_fault_tolerance}, we discuss how \cm handles failure scenarios. By removing state persistence from an invocation's critical path, \cm increases scheduling throughput, as we will show in \S\ref{sec:eval:microbenchmark}.}

\begin{table}[b]
  \begin{center}
    \centering
    \begin{tabularx}{\linewidth}{xyx}
         \hline
         Abstraction & Associated State & Persisted \\ 
         \hline 
         \hline
         
         \multirow{5}{*}{Function} & Name & \multicolumn{1}{c}{\checkmark} \\
         & Image URL & \multicolumn{1}{c}{\checkmark} \\
         & Port to expose & \multicolumn{1}{c}{\checkmark} \\
         & Scheduling configuration & \multicolumn{1}{c}{\checkmark} \\
         & Scheduling metrics & \\
         \hline
        
         \multirow{4}{*}{Sandbox} & Name & \\
         & IP address & \\
         & Port on worker node & \\
         & Worker node ID & \\
         \hline

         \multirow{2}{*}{DataPlane} & IP address & \multicolumn{1}{c}{\checkmark} \\ 
         & Port & \multicolumn{1}{c}{\checkmark} \\ 
         \hline
        
         \multirow{3}{*}{WorkerNode} & Name & \multicolumn{1}{c}{\checkmark} \\
         & IP address & \multicolumn{1}{c}{\checkmark} \\
         & Port & \multicolumn{1}{c}{\checkmark} \\
         \hline
        
    \end{tabularx}
    \caption{
        \cm's key cluster management abstractions and their associated state maintained by the control plane. 
    }
    \label{table:dirigent_abstractions}
  \end{center}
\end{table}

\textbf{Monolithic control and data planes.} \cm centralizes the functionality for creating and managing sandboxes into a monolithic control plane and the functionality for routing, throttling, and buffering function invocations into a monolithic data plane. 
\cm's monolithic architecture contrasts with systems like Knative and OpenWhisk, which inherit the microservice architecture of K8s where multiple components run as separate services and communicate via RPCs. \cm's monolithic control and data planes allow simpler deployment and management, fewer leader elections, and faster recovery time on crashes. In \cm's control plane, modules such as the state manager, health monitor, autoscaler, and placer exchange information through fast in-memory channels and atomic primitives. The monolithic data plane allows \cm to minimize infrastructure tax on warm starts, compared to \knative's approach of deploying separate Queue-Proxy sidecars per function sandbox for request buffering and throttling. 
Abolishing sidecars leads to faster sandbox startup time, better monitoring over invocations from data planes, less resource usage, and a shorter invocation critical path.
However, we decided to separate the control and data planes, such that we can scale data planes independently based on the warm invocation load while maintaining stable control plane performance for cold starts. 

\subsection{Life of a Request}\label{design:life-of-req}

We now describe how a function invocation traverses the \cm system in Figure~\ref{fig:dirigent_system_diagram}.
A function invocation arrives in \cm through the front-end load balancer (LB) and reverse proxy. If there is a sandbox to handle the invocation (i.e., a \emph{warm start}), the data plane picks a sandbox that will execute the invocation, ensures the sandbox has an available processing slot, and proxies the request to the worker node. If no sandboxes are available to process a request when it arrives (i.e., \emph{cold start}), the invocation waits in a data plane's request queue until at least one sandbox becomes available. The data plane periodically sends autoscaling metrics to the control plane. The autoscaler in the control plane determines the number of sandboxes needed to serve the current traffic. When a new sandbox needs to be created, the placer chooses and notifies the worker node that should spin up the new sandbox. Once a sandbox is created, the worker daemon issues health probes to ensure the sandbox is booted and ready to handle the traffic. After the sandbox passes a health probe, the worker daemon notifies the control plane, which then broadcasts endpoint updates to data plane components. The data plane dequeues the request and handles it as a warm start. Requests leave the system in the reverse direction and pass through the same data plane to reach the client.

\textbf{Synchronous vs. asynchronous invocations.} 
\cm supports both operation modes. Users specify the mode in the request header and submit the request as described above. Asynchronous calls pass through an additional queue between the front-end load balancer and the reverse proxy which submits requests and monitors invocation status and can be configured to re-invoke functions on timeouts. In this paper, we focus on synchronous calls as asynchronous ones are not supported by all FaaS platforms (e.g., Knative).
\subsection{Fault Tolerance}
\label{sec:system_design_fault_tolerance}

We discuss how \cm handles component-level and request-level fault tolerance. Furthermore, since \cm aims to minimize state persistence (\S\ref{sec:design_principles}), particularly on the critical path of invocations, we elaborate on how \cm matches the fault tolerance guarantees of today's FaaS platforms.

\subsubsection{Component-level fault tolerance}
\label{sec:component_level_fault_tolerance}

These failures occur because \cm's component(s) or the physical machines running them crash. \lazar{\cm leverages replication to recover components quickly, implements a restart policy on component failure, and ensures that requests arriving after any component failure are correctly executed.}

\textbf{Control plane fault tolerance.} \lazar{For high availability (HA), \cm runs multiple control plane components. One control plane component is the leader that serves requests, while others are on standby. Each control plane component runs an instance of the replicated cluster state database. New sandboxes cannot be spawned while the control plane leader is down. However, warm functions remain unaffected, provided the data plane does not crash. The control plane recovers by electing a new leader followed by fetching all \textit{DataPlane} and \textit{WorkerNode} objects from the persistent storage to re-establish connections with the cluster components. The control plane then retrieves \textit{Function} objects from the database and updates data plane caches. At this point, the control plane can serve new requests. The scale of all deployed functions in the control plane's internal data structures is zero, although, in a scenario where only the control plane crashes, worker nodes still run the sandboxes, i.e., the actual scale is greater than zero. Hence, worker nodes provide the control plane with a list of sandboxes they run and the control plane merges this information asynchronously, as it arrives, to its internal data structures and notifies data planes of changes. 
\cm does not downscale recovered sandboxes for one autoscaling time window (60s by default), since the autoscaling metrics, which were lost on failure, take time to repopulate.}

\textbf{Data plane fault tolerance.} The data plane is replicated. Each replica is active and operates independently. When a data plane component fails, it recovers by re-establishing a connection with the control plane and pulling the list of registered functions and sandboxes in the cluster.

\textbf{Worker node fault tolerance.} The worker node is considered healthy and schedulable as long as the control plane receives periodic heartbeats from it. Once the control plane detects no heartbeats, it notifies data plane components not to route requests to sandboxes on the affected worker node. The control plane re-runs autoscaling to spin up sandboxes somewhere else. The worker node continuously monitors sandbox processes and notifies the control plane of crashes.

\lazar{\textbf{Multi-component fault tolerance.}
\cm can tolerate failure of multiple components of different types, each of which individually recovers as described above. \cm cluster is operational as soon as at least 1 control plane, 1 data plane, and 1 worker node are available. In the worst failure scenario, when the control plane, data plane, and all worker nodes fail, the cluster after recovery will be equivalent to the cluster where all functions have zero sandboxes running. Since sandbox creation in \cm is quick (see \S\ref{sec:evaluation:cold_start_microbenchmark}), the cluster will converge to the state for serving the current traffic demand, while invocations will experience a slowdown during the convergence period. However, \cm does not guarantee the exact sandbox count as before the failure, nor that sandboxes will be assigned the same IP addresses or placement as before the failure.}

\lazar{\textbf{State consistency.} 
Data plane components operate on information from their internal caches, which can become stale if the control plane experiences longer downtime. In such scenarios, a data plane can load balance requests to a non-existing sandbox. \cm favors availability over consistency, similar to many production-grade load balancers~\cite{envoy_proxy:inconsistency}.}

\subsubsection{Request-level fault tolerance} 
\label{sec:system_design:request_level_ft}
Cluster manager component failures may lead to invocation failures. For example, if a worker node fails, all invocations executing on that node will also fail. If a data plane fails, all inflight requests in that data plane will be terminated, as connections to clients are lost. \cm provides no request-level fault tolerance guarantees for synchronous invocations, which is also the case with the \knative, OpenWhisk~\cite{openwhisk:invocation_level_guarantees}, and commercial FaaS platforms such as AWS Lambda and Azure Functions~\cite{aws:invocation_level_guarantees,azure:invocation_level_guarantees,gcf:invocation_level_guarantees}. 
\lazar{For synchronous requests, these systems rely on the user to re-invoke functions.} 
\lazar{For asynchronous requests, \cm provides at-least-once guarantees, through request persistence and a retry policy. \cm can serve as a basis for providing stronger request-level guarantees~\cite{boki,beldi,halfmoon}. Data plane and worker nodes can also be extended to support workflow orchestration, function checkpointing, and transactions~\cite{apiary}.}

\section{Implementation and Limitations}
\label{sec:implementation_details}

We implement \cm in approximately 11.3K lines of Go code. Communication between system components shown in Figure~\ref{fig:dirigent_system_diagram} happens via gRPC calls that are invokable at any time, rather than through periodic heartbeats like in Mesos and YARN~\cite{hindman:mesos,vavilapalli:yarn}. 
\cm uses RAFT~\cite{ongaro:raft} for control plane leader election and relies on \texttt{systemd} to monitor \cm component health and restart a failed process. 
\cm uses Redis~\cite{redis} to persist the system state. We collocate a Redis replica with each control plane component replica. When a control plane leader changes, the Redis master also changes.

\textbf{Concurrency.} System components use readers-writer locks for all critical sections with a transactional state update. On hot-paths, we use lock-free data structures where possible. The communication between different control plane modules such as the placer and autoscaler uses Go channels.

\textbf{Worker node software stack.} We implement \cm with two different sandbox runtimes: containerd~\cite{containerd} and Firecracker~\cite{agache:firecracker, brooker:firecracker_snapshots} with and without microVM snapshots. Integrating additional sandbox runtimes only involves extending a three-call interface. Each worker node maintains a local container image and snapshot cache to reduce image pulling. Because of Linux network stack performance issues on parallel network interface creations~\cite{mohan:agile_cold_starts,particle}, each worker node maintains a pool of pre-created recyclable network configurations along with pre-configured iptables rules to allow quick network allocation to a newly created sandbox.

\textbf{Scheduling policies.}
\cm implements and uses Knative's default scheduling policies across all three scheduling dimensions (autoscaling, placement, and load balancing). The autoscaling algorithm scales the number of sandboxes per function based on the number of in-flight requests for each function~\cite{knative:autoscaling_policy}. 
The placement policy favors nodes with the least utilized resources while aiming to balance resource utilization across CPU and memory~\cite{kubernetes:placement_policy}. 
\cm supports Hermod~\cite{kaffes:hermod} and CH-RLU~\cite{fuerst:Locality-Aware-Load-Balancing} scheduling policies, though they are unused in our evaluation (\S\ref{sec:evaluation}) to ensure a fair comparison to Knative. 
The load-balancing algorithm forwards invocations to least-loaded sandboxes~\cite{knative:load_balancing_policy}.
\lazar{The front-end load balancer steers invocations based on function ID hash, which ensures all invocations of a particular function end up on the same data plane component and allows centralized tracking of the number of in-flight requests for each function.} 
Implementing new scheduling policies and metrics involves extending the relevant Go interfaces in the control plane (for autoscaling and placement policies) and in the data plane (for load-balancing policies), recompiling, and redeploying \cm. \knative also requires recompilation, repackaging, and redeployment of its autoscaling, load-balancing, or placement service containers to add new policies and metrics. 

\lazar{\textbf{Sandbox teardown.} The control plane runs an asynchronous autoscaling loop that issues sandbox teardown calls to worker nodes, based on the inflight request count in the cluster. On worker nodes, such calls trigger sandbox termination, a process that dismantles the file system, network interfaces, and cgroups structures associated with the sandbox.}

\textbf{Operations and monitoring.} \cm components expose global and per-function metrics (e.g., the number of inflight requests, queue depth, and number of successful invocations) via HTTP, similar to \knative. \cm is equipped with logging infrastructure that reports important events in the cluster, eases debugging, and can be used to break down end-to-end function latency. \cm's logging and monitoring infrastructure provides a foundation for building fine-grain resource accounting and billing services.

\textbf{Limitations.} \cm does not currently support function versioning and partial traffic steering to different function versions, which is supported in \knative. This can be implemented in \cm by extending \textit{Function} and \textit{Sandbox} abstractions with a version number and by adding a versioning-aware load-balancing policy in the data plane. Cluster manager features like QoS support and remote log fetching are not yet integrated into \cm but can be added. 
We emphasize that \cm is an alternative to FaaS cluster managers. It is not intended as a replacement for a general-purpose cluster manager as it does not support naming/discovery services for coordination between sandboxes or provide strict guarantees for state reconstruction upon failures as K8s.

\section{Evaluation}
\label{sec:evaluation}

We evaluate \cm to answer the following key questions:

\begin{itemize}
    \item What is the throughput of \cm's control plane, i.e., what is the system's peak sandbox creation rate? 
    \item What is the \cm's data plane throughput, i.e., how many warm requests can \cm serve per second? 
    \item How does \cm improve end-to-end function latency and cluster resource utilization for FaaS production workload compared to state-of-the-art systems?
    \item How effectively does \cm handle control plane, data plane, and worker node failure scenarios? 
\end{itemize}

\subsection{Experimental Methodology}

\textbf{Baselines.} We compare \cm to two open-source K8s-based FaaS platforms: Knative~\cite{knative} and OpenWhisk~\cite{openwhisk}. We briefly experimented with OpenFaaS~\cite{openfaas} as another K8s-based baseline, but we found that the community version is not competitive as it only supports up to 15 functions and lacks critical features like scale-to-zero and concurrency throttling. We compare \cm's end-to-end performance to a state-of-the-art commercial platform, AWS Lambda~\cite{aws:lambda}.

\textbf{Hardware setup.} We run \cm and the open-source baseline systems on a 100-node xl170 Cloudlab cluster~\cite{cloudlab}.
Each node is an Intel Xeon E5-2640 v4 @ 2.4~GHz CPU with 10 physical cores, 64GB of DRAM, and an Intel DC S3520 SSD. All nodes run Ubuntu 20.04. Nodes are connected in groups of 40 machines with 25~Gbps links to Mellanox 2410 leaf switches and groups connect to a Mellanox 2700 spine switch with 100~Gbps links. For AWS Lambda experiments, we register functions in the us-east-1 region and invoke functions from T3 EC2 instances in the same region.

\textbf{Software setup.} We run Knative v1.13.1~\cite{knative} with Istio v1.20.2~\cite{istio} and OpenWhisk v1.0.1~\cite{openwhisk}. Both baselines run on Kubernetes v1.29.1~\cite{kubernetes}.
We use containerd v1.6.18~\cite{containerd} as the sandbox manager. \cm also supports snapshot-enabled Firecracker v1.7.0~\cite{agache:firecracker} sandboxes. Firecracker microVMs run Linux kernel v4.14. For the persistent data store, \cm uses Redis v7.2.0~\cite{redis} in append-only mode with fsync enabled at each query. We use HAProxy v2.4.24~\cite{haproxy} with keepalived v2.2.8~\cite{keepalived} as a highly-available front-end load balancer. We configure sandboxes to handle only one request at a time, similar to commercial cloud offerings~\cite{aws:sandbox_concurrency,gcf:invocation_level_guarantees}. We employ the same scheduling policies in \knative and \cm (\S\ref{sec:implementation_details}), and prefetch container images and VM snapshots on each worker node. We do container image prefetching in AWS Lambda experiments with technique from~\cite{brooker:firecracker_snapshots}.

\lazar{In both Knative and \cm experiments, we run the control plane in high-availability (HA) mode with 3 replicas, each running on a dedicated node. Also, we run 3 data plane replicas on separate nodes. We co-locate the front-end load balancer with the data planes and run the InVitro~\cite{ustiugov:in_vitro} load generator on a separate machine in the cluster.}

\subsection{Microbenchmarks}
\label{sec:eval:microbenchmark}

We analyze cluster manager latency, peak throughput, and scalability by invoking hello-world functions. We run cold start microbenchmarks to stress test the control plane and warm start microbenchmarks to stress test the data plane.  

\subsubsection{Cold Start Performance}

\begin{figure}
    \centering
    \includegraphics[trim={0 0.75cm 0 0},width=\linewidth]{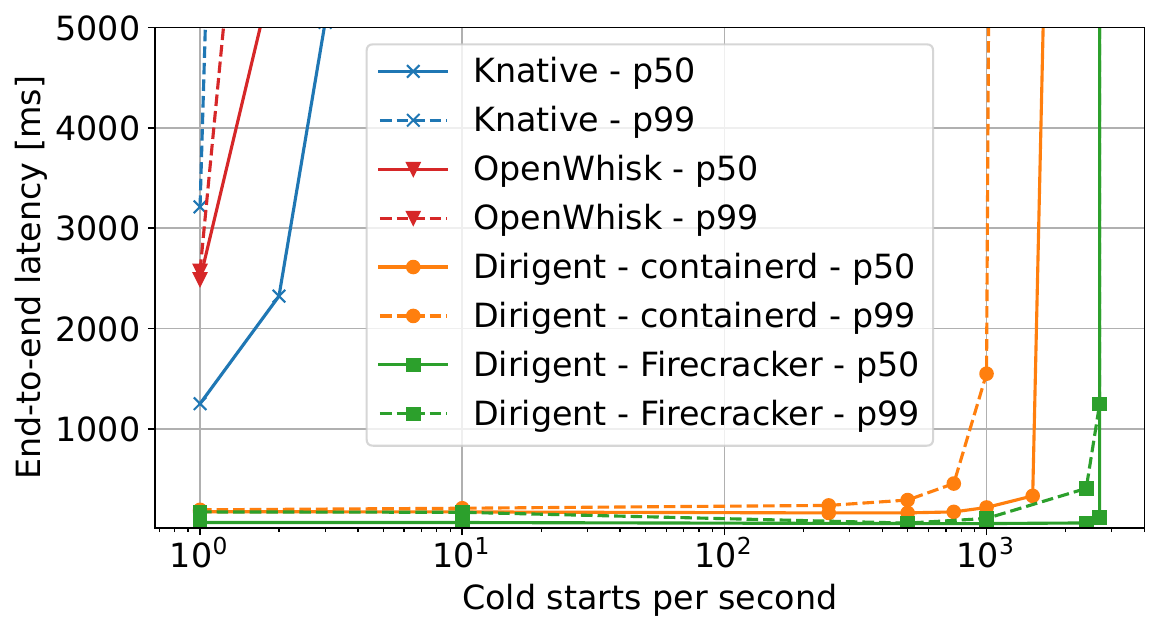}
    \caption{Cold start performance.}
    \label{fig:system_performance_comparison}
\end{figure}

\textbf{\\Peak sandbox creation throughput.} 
\label{sec:evaluation:cold_start_microbenchmark}
Figure~\ref{fig:system_performance_comparison} shows the p50 and p99 end-to-end latency as we sweep the number of cold start invocations per second in the 93 worker-node cluster. \cm sandbox creation throughput with containerd saturates at 1750 cold starts per second. The bottleneck is not the \cm control plane but kernel lock contention during sandbox creation, network interface configuration, and iptables rule updates on containerd worker nodes. To saturate the \cm control plane, we optimize the worker node software stack by running functions in Firecracker microVMs booted from snapshots. \cm with Firecracker microVMs achieves a peak throughput of 2500 cold starts per second. At this load, the \cm control plane CPU utilization is still only 55\%
and access congestion on shared data structures used for autoscaling becomes the bottleneck. 
In contrast, cold start latency with \knative and OpenWhisk saturates at significantly lower load (below 2 cold starts per second!), due to high CPU utilization on the K8s API Server which is processing many RPCs from controller components and serializing large volumes of data for state updates to the etcd database. 
Note that compared to the experiment in Figure~\ref{fig:concurrent_sandbox_creation_sweep}, where we invoked bursts of specific size and reported the p50 latency for invocations in that burst, here we invoke functions at a steady rate.
Overall, \cm enables 1250$\times$ higher sandbox creation throughput than the K8s-based cluster managers. This is critical as FaaS clusters in production experience bursts in which thousands of sandboxes per second must be created (recall Figure~\ref{fig:invocation-load}).

\textbf{Cold start latency breakdown.} Figure~\ref{fig:system_performance_comparison} also shows that \cm's cold start latency is lower than K8s-based systems \textit{even at low load} (e.g., 1 cold start per second). 
We analyze the breakdown of unloaded cold start latency in \knative and \cm.
\knative is slow at booting new sandboxes ($\sim$400~ms) 
since in addition to the user container, it creates a queue-proxy sidecar container on the worker node for each user function container. The sidecar buffers requests to the user container. These two containers are created sequentially and need to pass the readiness probe checks, which we find takes $\sim$500~ms 
after both containers are created. In contrast, \cm buffers requests in per-function queues in data plane nodes and therefore does not need to boot sidecars on worker nodes in the critical path. This significantly reduces sandbox creation and readiness wait latency. \cm also has lower control plane latency due to minimal state updates on the critical path of sandbox creation. \cm with Firecracker snapshot microVMs further reduces unloaded cold start latency as it reduces sandbox creation and network configuration latency on worker nodes.  

\textbf{\cm optimization breakdown.} To understand which aspects of \cm's design contribute most to performance benefits, we repeat the cold start throughput sweep experiment with a modified version of \cm that persists all state in Table~\ref{table:dirigent_abstractions}, including sandbox state. Persisting sandbox state in the control plane introduces a write to persistent storage on the critical path for cold starts, which decreases \cm's peak cold start throughput to 1000 cold starts per second, and p99 latency surges at 500 cold starts per second. This confirms that avoiding persistent state updates on the critical path of cold start requests is a performance-critical design decision. In \S\ref{sec:evaluation:fault_tolerance} we will show this design decision does not degrade failure recovery times, as \cm can still reconstruct sandbox state efficiently from worker nodes in case of control plane failures. We also confirm that simply fusing K8s components (which avoids RPCs between controllers) is not sufficient to eliminate performance issues in K8s-based cluster managers. We deploy Knative on top of K3s~\cite{k3s}, which is a monolithic implementation of K8s within a single process. We observe only marginally higher peak cold start throughput than \knative on K8s, indicating that the state management and state persistence design decisions are much more performance-critical than the monolithic control plane. However, \cm's monolithic control plane is still useful as it simplifies the system design and deployment. 
\lazar{Finally, reducing the volume of state that \cm manages is also a performance-critical design decision since it avoids saturating the CPU with data structure serialization tasks, which we saw in  \S\ref{background:k8s-mismatch} limit the scheduling throughput of K8s. } 

\subsubsection{Warm Start Performance}

\begin{figure}
    \centering
    \includegraphics[trim={0 0.75cm 0 0},width=\linewidth]{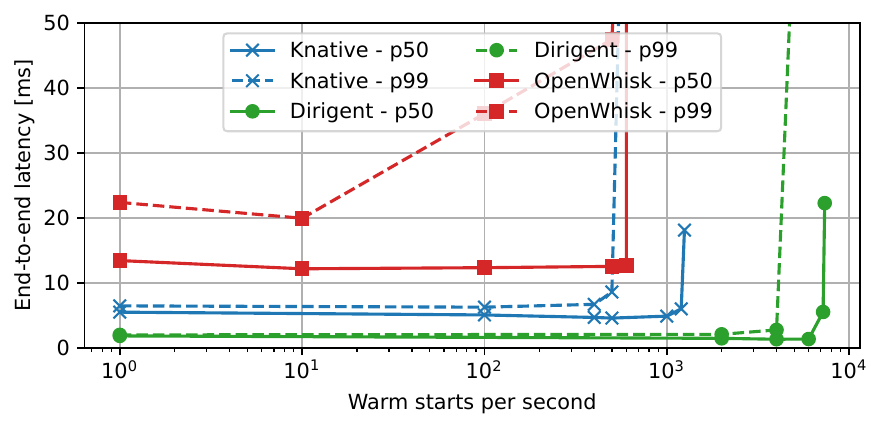}
    \caption{Warm start performance.}
    \label{fig:steady_load}
\end{figure}

To stress-test the cluster manager data plane, we now consider only warm starts, i.e., invocations for which a sandbox is already available in the cluster and the control plane is not on the critical path. 
Figure~\ref{fig:steady_load} shows the p50 and p99 end-to-end latency as we sweep warm start throughput. \cm can sustain 4000 warm invocations per second with a p50 latency of 1.4~ms and a p99 latency of 2.5~ms. 
The components that contribute to the warm start latency are the front-end load balancer, proxy service, request throttler on data plane nodes, and iptables NAT on worker nodes. 
At the peak warm start throughput, \cm cannot accept any new requests since the machine runs out of ports.
In contrast, \knative achieves a peak throughput of only 1200 warm starts per second with a p50 latency of 7~ms, as the activator and queue-proxy components in Knative add delays. OpenWhisk's high latency originates from its architecture, where Apache Kafka~\cite{apache:kafka} and CouchDB~\cite{apache:couchdb} are on each request's critical path~\cite{fuerst:iluvatar}.

\subsubsection{Scalability}
\label{sec:evaluation:worker_node_scalability}

We explore how cold start throughput scales as we increase the number of worker nodes in the cluster.
\knative claims to support clusters of up to 5K nodes~\cite{kubernetes:scalability_limit}. 
Since we do not have access to thousands of nodes, we run multiple worker daemons per machine on our 100-node cluster. Each worker daemon sends heartbeats to the control plane and sleeps for 40~ms upon receiving a sandbox creation request, which corresponds to the p50 Firecracker microVM creation time from snapshots. We find \cm latency and peak throughput match the results in Figure~\ref{fig:system_performance_comparison} when cold starts are distributed across up to 2500 worker nodes. With more worker nodes, throughput starts to degrade (e.g., with 5000 workers, \cm supports up to 2000 cold starts per second) due to contention on shared data structures for monitoring sandbox health in response to heartbeats. 

\lazar{While we have so far shown the scalability of a single \cm cluster, \cm can further scale by dividing big clusters into smaller sub-clusters, analogous to Borg cells~\cite{borg}. In such a deployment, each sub-cluster runs its own control and data plane components, while a front-end sharding system~\cite{adya:slicer,lee:shard_manager} steers invocations to sub-clusters.}

\subsubsection{Function Registration Performance}

Before a user can invoke a function, they must first register the function. Although registration is only done once per function, fast registration is important for quickly deploying applications with many functions. \knative takes roughly 18 minutes, whereas \cm takes 1 second.
In \knative, it takes $\sim$770~ms to register a single function in an empty cluster, but this latency grows the more functions there are in the system. This is because \knative ascribes multiple abstractions to each function on registration (e.g., routes, revisions, services) and synchronizes ingress controllers. In contrast, registering a function in \cm takes 2~ms on average, as it only involves persisting function specification into the database and propagating metadata to data plane components.

\subsection{End-to-End Performance on Azure Trace}
\label{sec:eval:azure}

We now measure end-to-end performance on a FaaS production workload trace from Microsoft Azure~\cite{shahrad:serverless} that contains 70K functions invoked over two weeks. We use InVitro~\cite{ustiugov:in_vitro} to obtain a representative trace sample that can run on our 100-node cluster. We extract a 30-minute time window starting in the middle of the trace (8th hour of day 6) and sample 500 functions trace with 168K invocations. We also test \cm with a larger trace containing 4K functions and 3.33M invocations. Functions execute the SQRTSD x86 instruction for a number of iterations derived from the function execution time distribution in the trace. We run experiments for 30 minutes and discard the first 10 minutes as a warm-up.

We measure scheduling latency and per-function slowdown. Slowdown is the end-to-end latency of the invocation in the FaaS cluster divided by the function's execution time on a dedicated worker node with no cluster scheduling overhead. Since the execution times of different functions in the trace can vary by orders of magnitude, we group by function and report the geometric mean slowdown per function. 
We also evaluate resource efficiency by measuring cluster CPU and memory usage. Since OpenWhisk performance is worse than \knative for both cold and warm starts in \S\ref{sec:eval:microbenchmark}, we do not include it here, but we compare to AWS Lambda.

\begin{figure}
    \centering
    \includegraphics[trim={0 0.75cm 0 0},width=\linewidth]{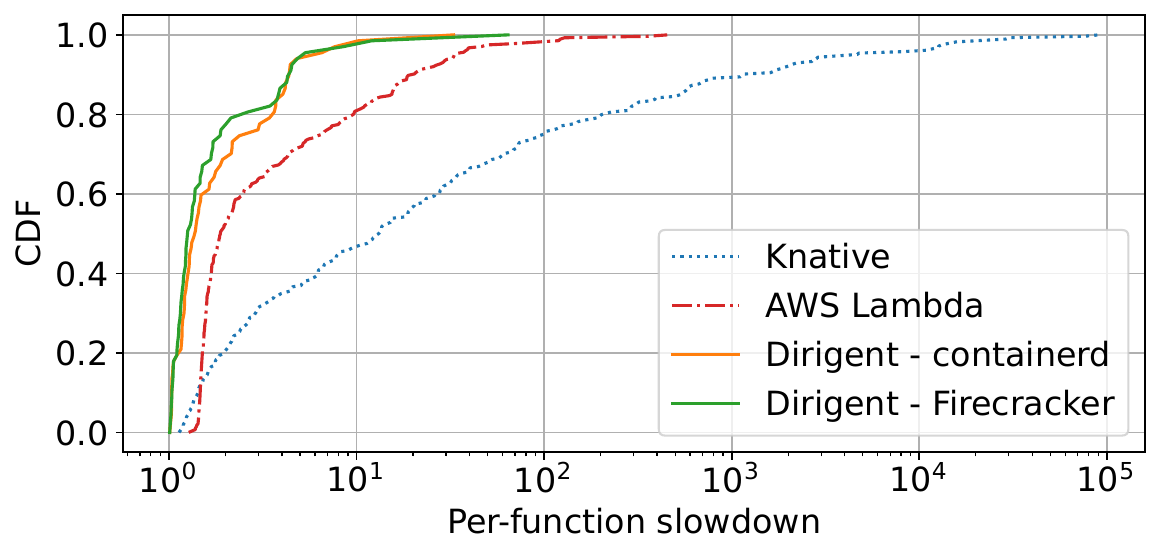}
    \caption{Per-function slowdown CDF for Azure 500 trace.}
    \label{fig:e2e_latency_azure_trace}
\end{figure}

\begin{figure}
    \centering
    \includegraphics[trim={0 0.75cm 0 0},width=\linewidth]{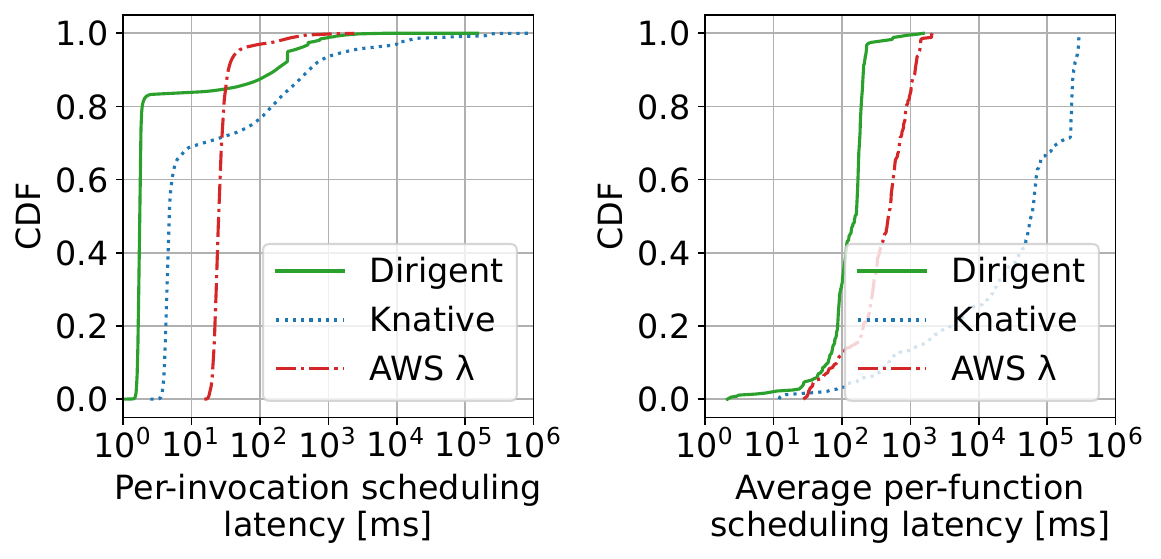}
    \caption{Scheduling latency for Azure 500 function trace.}
    \label{fig:eval:e2e_scheduling_latency}
\end{figure}

\textbf{Function latency analysis.} Figure~\ref{fig:e2e_latency_azure_trace} shows \cm significantly reduces per-function slowdown compared to state-of-the-art systems. While the median function slowdown is 1.87 in AWS Lambda and 13.2 in \knative, it is only 1.38 with \cm. \cm especially reduces scheduling overheads at the tail, i.e., reduces p99 function slowdown by 6.89$\times$ compared to AWS Lambda and by over three orders of magnitude compared to \knative. While slowdown quantifies the impact the cluster manager has on end-to-end latency (which also depends on the function's execution time), Figure~\ref{fig:eval:e2e_scheduling_latency} shows the raw scheduling latency CDFs for the same experiment, both per-invocation and per-function average scheduling latency. Note the log scale. \cm reduces the median and p99 per-function scheduling delay by 3.07$\times$ and 2.79$\times$ compared to AWS Lambda, respectively. \cm reduces the p99 per-function scheduling delay by 403$\times$ compared to \knative.

The functions that experience the highest slowdown in \cm are those with the shortest execution time (i.e., below 10~ms) as these functions are the most sensitive to scheduling overheads and sandbox creation delays. Meanwhile, the functions with the highest slowdown in \knative and AWS Lambda experiments are predominantly functions whose individual invocations are greatly spread out over time but occur during times in the trace when the cluster experiences the most cold starts. We find some functions in the trace are repeatedly invoked in unison (due to timer-based invocation triggers~\cite{shahrad:serverless}) with long periods, resulting in large cold start bursts in the cluster. These bursts lead to high scheduling latency in AWS Lambda and \knative, whereas \cm handles much higher cold start throughput. 
For the Azure 500 function trace experiment, \knative's median per-invocation scheduling latency is 4.67~ms and 59.59~s at the 99th percentile. In contrast, \cm's median scheduling latency is 1.74~ms and 1.13~s at the 99th percentile. \cm with Firecracker has a bit longer per-function slowdown tail as some functions are never invoked during the warm-up period and depend on the disk for snapshot restoration.

\textbf{Sandbox creation count.} \cm creates fewer sandboxes throughout the experiment even though it uses the same autoscaling algorithm and metrics as \knative. During the experiment, \knative spawned 2930 sandboxes, whereas \cm created only 713 sandboxes for the same workload trace. To understand this discrepancy, we need to delve into the functioning of the Knative autoscaling algorithm. Knative's autoscaler monitors the number of inflight requests, which includes both those actively being processed within pods and those queued. The desired number of pods is directly proportional to the inflight request count. Intuitively, when a queue forms, the autoscaler initiates new pod creations proportionally to the queue length. However, due to a lengthy scale-up delay within \knative, the queue continues to grow during the scale-up process, prompting the creation of even more pods. In contrast, \cm exhibits a more responsive behavior. When a queue starts to form, the Knative autoscaling algorithm starts creating pods, and \cm promptly scales the number of ready pods to the desired state of the autoscaler, leading to a near-immediate depletion of the queue. This swift response translates to a significantly reduced number of pods being provisioned overall.

\textbf{Resource utilization.} We observe \cm control plane node only uses 3\% of CPU cycles on average, whereas in \knative, the CPU is consistently above 75\% utilized struggling to handle cold start bursts. \cm provides higher scheduling performance while consuming fewer CPU resources for the control plane than \knative. Memory on worker nodes in \knative and \cm is utilized 4.62\% and 3.1\%, respectively.

\textbf{Larger trace.}
While the sampled Azure trace with 500 functions is the biggest trace we can run with \knative before we start observing high invocation failure rates due to timeouts, this trace can not saturate the same hardware cluster orchestrated by \cm. Hence, we run a larger Azure trace sample with 4000 functions and 3.33M invocations. We compare \cm to AWS Lambda. With this trace, \cm utilizes 70\% of CPU resources on worker nodes and achieves p50 and p99 slowdowns of 2.14 and 15.4, respectively. On the other hand, AWS Lambda's p50 and p99 slowdowns are 70 and 11631, respectively. Finally, \cm experiences a negligible invocation failure rate, while in the AWS Lambda, 33\% of invocations experience timeout.

\subsection{Fault Tolerance}
\label{sec:evaluation:fault_tolerance}

We now analyze the impact of component failures. We measure average function invocation slowdown over time for the Azure 500-function workload, while triggering
failures.

\begin{figure}
    \centering
    \includegraphics[trim={0 0.75cm 0 0},width=\linewidth]{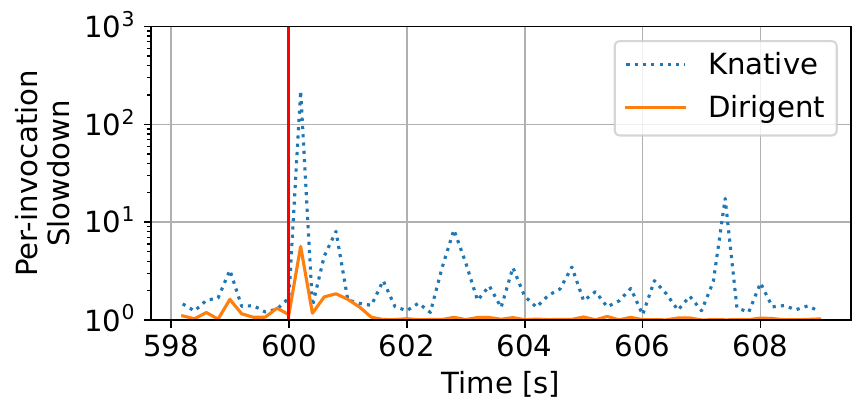}
    \caption{Control plane fault tolerance. The vertical red line shows when the failure occurs.}
    \label{fig:fault_tolerance:control_plane}
\end{figure}

\textbf{Control plane failure.} 
Figure \ref{fig:fault_tolerance:control_plane}, shows how the slowdown of function invocations varies over time before and after we fail the control plane leader for \cm and \knative. A control plane failure impacts performance by adding a queuing delay for cold starts. Such invocations remain buffered in the data plane until the control plane becomes operational to schedule a sandbox creation or until a busy sandbox related to that function on some worker node becomes idle.
\cm achieves a lower per-invocation slowdown for invocations issued at the moment of failure and stabilizes the slowdown quicker than \knative. 
\lazar{The performance improvements of \cm stem from the fast control plane recovery mechanism that takes 10ms to detect a control plane leader failure, elect a new leader, retrieve recovery-relevant information from the DB, and synchronize data planes.}
In \knative, it can take several seconds until each control plane microservice recovers and the control plane starts serving new requests.

\textbf{Data plane failure.} 
When a data plane fails, all inflight requests associated with that data plane also fail, as clients' connections are terminated. We fail one data plane replica and monitor the invocation failure rate. In \cm, we observe it takes 2s for the invocation failure rate to stabilize at zero after a data plane failure.
\lazar{The recovery time includes failure detection, restarting the systemd service, re-connecting with the control plane, synchronizing data plane caches, re-configuring the front-end load balancer, and depleting the load balancer queue.}
In \knative, whose data plane is not a monolith as in \cm, we measured it took 15s for the data plane to recover. We observe Istio Ingress Gateway dominates the recovery time, as the slowest component to restart.

\textbf{Worker daemon failure.} When the worker daemon on a node fails, the worker can no longer respond to any control plane commands, including starting or tearing down sandboxes. This leads to a higher slowdown on cold invocations, while warm invocations remain affected. We failed 47 out of 93 worker daemons in the cluster while monitoring the slowdown of functions invoked during worker downtime. \cm achieves a peak per-invocation slowdown of 2.7, which is 10$\times$ lower than \knative, as \cm can efficiently create new sandboxes on non-affected nodes and because it has shorter worker daemon recovery time.

\textbf{Concurrent component failures.} 
\cm remains operational as long as one control plane replica is elected as a leader and at least one data plane is operational. In case of concurrent component failures, the recovery time will be dominated by the slowest component to recover, as components can recover in parallel.

\section{Future Directions}\label{sec:discussion}

By enabling orders of magnitude higher sandbox creation throughput compared to existing platforms, \cm provides a foundation for future FaaS system research. We are currently exploring how \cm's design generalizes to scheduling function workflows by extending \cm data plane components to serve as workflow orchestrators. We also aim to explore the performance trade-offs related to providing stricter request-level fault tolerance guarantees, such as at-least-once or exactly-once~\cite{shim,boki,beldi,halfmoon,apiary}, and quantifying their cost at scale. We plan to integrate additional sandbox runtimes~\cite{dandelion} and scheduling policies. Another future direction involves exploring caching techniques for sandbox images and snapshots at scale~\cite{brooker:firecracker_snapshots}.

\section{Conclusion}\label{sec:conclusion}

\cm is a new customized cluster manager for serverless. In contrast to the state-of-the-art approach of building FaaS cluster managers on top of legacy cluster managers like Kubernetes, \cm presents a clean-slate system architecture, simple abstractions, and lightweight persistence for state management to eliminate the performance bottlenecks of K8s-based cluster managers in high-churn FaaS environments. 
We show that \cm can schedule 2500 sandboxes per second at low latency, which is 1250$\times$ more than Knative. \cm achieves 6.89$\times$ lower 99th percentile per-function slowdown and 403$\times$ lower 99th percentile per-function scheduling latency compared to \knative on a production Azure trace while maintaining 25$\times$ lower control plane CPU utilization on average. \cm also improves recovery times from component failures compared to \knative.

\begin{acks} 
\lazar{We thank Rodrigo Fonseca, Lalith Suresh, Timothy Roscoe, Michael Wawrzoniak, Patrick Stuedi, and Malte Schwarzkopf for their valuable feedback. We also thank our anonymous shepherd and reviewers for their helpful comments and suggestions. Thank you to Luka Simi\'{c} and the anonymous artifact evaluators for verifying our experiment results.}
\end{acks}

\balance
\bibliographystyle{acm}
\bibliography{ref}

\end{document}